# AGAINST FREE WILL IN THE CONTEMPORARY NATURAL SCIENCES[1]

Martín López-Corredoira[2]
Instituto de Astrofísica de Canarias
La Laguna, Tenerife (Spain)

**Abstract:** The claim of the freedom of the will (understood as an individual who is transcendent to Nature) in the name of XXth century scientific knowledge, against the perspective of XVIIIth-XIXth century scientific materialism, is analysed and refuted in the present paper. The hypothesis of reductionism finds no obstacle within contemporary natural sciences. Determinism in classical physics is irrefutable, unless classical physics is itself refuted. From quantum mechanics, some authors argue that free will is possible because there is an ontological indeterminism in the natural laws, and that the mind is responsible for the wave function collapse of matter, which leads to a choice among the different possibilities for the body. However, here I defend the opposite thesis because indeterminism does not imply free will, and because the considerations about an autonomous mind sending orders to the body is against neuroscience or evolutionary theories about human beings. The quantum theory of measurement can be interpreted without the intervention of human minds, but other fields of science cannot contemplate the mentalist scenario. A fatalistic or materialist view, which denies the possibility of a free will, makes much more sense in scientific terms.

## 1. Introduction

The negation of free will in the name of scientific knowledge is an issue, in my opinion, that was finally settled by many authors in the XVIIIth and XIXth centuries. Nevertheless, the XXth century brought forth many proponents of free will within the framework of the contemporary natural sciences, for which reason it would seem appropriate to revisit the question. In this essay it is my intention to argue against these libertarian positions. In view of the proliferation of works that misuse scientific terminology in order to tackle metaphysical subjects such as freedom, I take this opportunity to review the arguments against free will put forward by classical fatalistic philosophers such as Hobbes, Spinoza, and Schopenhauer, as well as to consider what modern science has brought to the topic. For this purpose, I shall deal with matters that

---

are currently in fashion in science today: quantum mechanics, the thermodynamics of irreversible processes, chaos, neuroscience, evolution, etc., and I examine the ideas of such authors as Bohr, Heisenberg, Boltzmann, Maxwell, Einstein, Prigogine, Penrose, Eccles, Crick, Popper, and a number of other scientists and philosophers of the last century. The main object of this essay is to defend the thesis that, contrary to the claims by some contemporary authors (e.g., Penrose and Prigogine), XXth century science did not in fact provide new insights into free will or dualism-mentalism.

The contemporary perspective, far from presenting a united front, either historically or at present, follows many widely divergent lines of argument, even to the point of being mutually contradictory. We cannot speak of contemporary thought as a unified structure, or of any current idea of a given concept, and even less may we refer to such a philosophical concept as freedom, concerning which there is not, there never has been, nor ever will easily be reached the slightest degree of consensus - even though, in principle, such consensus is not impossible. What I shall do is guide my arguments towards a certain way of seeing the world, which in passing I state as precisely the one I aim to defend, as it is usual to do. This discussion will be centred on the ambit of scientific thought. Even scientific thought is not monolithic, but it does at least limit considerably the number of possibilities. I mean "science", not in the wide sense that it is used today, which embraces almost any branch of human knowledge, including some of the humanities, in which the general sense of the term "scientific knowledge" refers to the independence of private opinion and the objective quest for a general consensus on that knowledge (Ziman, 1968) but in a fairly restrictive sense. I understand the term within the context of Comte (1842) classical positivism [*Cours de philosophie positive*], which was limited to the natural sciences. I shall refer specifically to physics, chemistry, biology, geology, astronomy, and other physical sciences. I shall not, therefore, refer to formal sciences such as logic or mathematics, or to the social sciences, such as sociology, or economics, which are regarded by Comte as sciences but outside of the ambit of the natural sciences. Neither do I include in my discussion psychology or what are generally denominated the cognitive sciences, which are far removed from what are classically regarded as science.

In reality, this essay is not intended as a proposal to use science to provide definitive solutions to problems in metaphysics, but rather simply to interpret the current scientific position *vis-à-vis* free will. Of course, there are other kinds of analysis apart from the interpretation of scientific results, but they are not the subject of this essay. I believe that the humanities, far from being pure sciences, have much more to contribute to this debate. There are lucid analyses of the place of man in the literary panorama, for example, in which, according to specific cases, the independence or otherwise of human beings from their environment is emphasized. There are great works of history that trace the causes that lead humanity to pursue given directions. In short, there are many intellectual disciplines that shed light on the problem. Those branches of philosophy at once far removed from the sciences but close to history – politics and sociology— also provide clarifying insights into signalling the repression of individuals immersed in a totalitarian society or their manipulation in the midst of the capitalist panorama. Such dark forces, related to the fatal quest for power that drive human beings are also related to the fate of human nature, which is part of the *fatum* of a mechanical universe, whose laws extend from the simplest systems to the complex mechanisms of social mechanics. All this is very interesting, but here I restrict my treatment exclusively to the natural sciences because such an analysis can be more precise and rational arguments of better clarity can be brought to bear. Moreover, this approach is sufficient by itself for an argument against free will.

The natural sciences allow knowledge to be gained of human nature. These sciences do not provide the correct solution to all problems, but knowledge at least of human nature is not proscribed to them. We may speak of human nature in scientific terms in the same way that we may consider other parts of nature; even where such investigation meets a barrier of understanding, science is still able to determine when we may speak of ordinary matter from the point of view of physics, or when it will become necessary to speak of other kinds of substances, behaviour, motions, etc., that elude ordinary laws. It is nevertheless in order to call into question the capacity of science to deal with these questions, and that is exactly what I propose to do here.

There is a certain consensus in the scientific world at large. Combining science and freedom, as is frequently done nowadays, is tantamount to asking oneself, as does Scott (1985) throughout his book *Atoms of the Living Flame*, how do our desires control atoms? That our bodies are made of atoms is not subject to the slightest doubt in contemporary science, and our acts produce the movement of these atoms. The only possibility of freedom for man is through the movement of the matter that constitutes his body through orders given by the autonomous self will of the individual rather than merely according to the laws of nature. That is what this essay is about.

"Men should know that from the brain, and only from the brain, stem all joy, pleasure, laughter and leisure, all grief, pain, dejection and lamentation," said Hippocrates two-and-a-half thousand years ago. This same affirmation could well have issued from the mouth of a materialist scientist of two or three centuries ago. Thus La Mettrie (1749) says:

> [I]f what my brain thinks is not part of that visceral organ, and if therefore not part of my body, why then, when I lie peacefully in my bed do I outline the plan of a work or pursue abstract reasoning, does my blood warm? (...) Because, in the final analysis, if the tension of my nerves, which constitutes pain, gives rise to the fever through which the spirit is disturbed involuntarily and if, reciprocally, the overworked spirit agitates the body and lights that fire of that consumption that carried off Bayle at such a young age, if such excitement urges me to want, forces me ardently to desire what a moment ago that which did not concern me in the least, if in their turn certain whims of the brain provoke certain urges and desires, why bother to duplicate what is evidently one?

La Mettrie was a physician and knew at first hand what the human body was and how its mechanisms worked. Also a physician was the XIXth century German thinker Büchner, who, in his book *Force and Matter* (Büchner, 1855, ch. 7) says:

> All that we have said regarding the relationship between force and matter leads us to conclude that rational and natural laws are always identical. What we call spirit, understanding, intelligence consist of natural forces, albeit combined in a particular manner such that, for their part, and like any other natural force, they can manifest themselves only in certain determined forms of matter. Being combined in organic life in an indefinitely complicated way and in particular forms, they produce effects that, at first sight, appear to us strange and inexplicable, whereas the processes and effects of the inorganic world are infinitely simpler and therefore easier to understand. Fundamentally, however, it remains the same matter, and experience continually teaches us that the laws of intelligence are the very laws of the world.

The fatalism of classical science can be summed up in an expression of Laplace (1814) inspired by Newtonian physics:

> An intelligence that could know at every moment all the forces that drive Nature, as well as the respective positions of all the beings that comprise it, that furthermore were sufficiently ample as to submit all such data to analysis could embrace in a single formula the movements of both the largest bodies in the Universe and those of the slightest atom; there would reign no uncertainty, and both the future and the past would present themselves before his eyes.

This quote from Laplace marks the starting point of this essay. Has the perspective of science changed noticeably from that of Laplace, La Mettrie, etc., with respect to the position of man in Nature? I basically distinguish among four types of answers to this question:

- Compatibilist: that the Laplacian Universe is not incompatible with a certain type of freedom. I reply to this in section 2.
- Non-reductionist: that the freedom of man is saved because man in not reducible to his constituent atoms. To be replied in section 3.
- That classical physics is not deterministic, so that Laplace's argument is flawed. I reply it in section 4.
- That there is an ontological indeterminism and subjectivism in the new quantum physics, such that, within this new representation of science, there is room for free will. My reply to this is in section 5.

## 2. Definition of freedom or free will

Before proceeding further, it is necessary to clarify the meaning of terms "freedom" (or "free will").

Freedom or free will: the spontaneous, unconditioned origin, without causal antecedents outside of us, from which we may consciously develop movements, choices, thoughts, sentiments, etc. It is the capacity to generate or be the origin of our movements, choices, etc., starting from nothing; that is to say, to be the first cause of what we may choose to develop. In other words, we posit a hypothetical personal capacity to choose our options and, therefore, our actions. This choice issues from our spontaneity, and external motivation cannot be what leads us to follow one path and reject another. The deliberative background of our actions is raised to the status of cause. Being free consists in possessing an inner source that is totally isolated from that Nature that produces "something" that releases our movements, choices, etc. That Nature is not the ego, and ego is not Nature. Phenomena exist in themselves and the identity of man, being something, is distinct from the identity of the phenomenon. The universality of phenomena must be excepted in the human being if the latter really wants to call itself free. To possess free will is to be something that is not contained in the Universe or governed by its laws, which precede our own existence. This meaning of freedom speaks to us of the relation man-Universe, of the possibility of removing ourselves from a cosmic order outside of ourselves. In the same way that we speak of certain communities having autonomy in certain areas of social legality with respect to the State within which it can pass its own laws, so may the human being be called free with regard to determined kinds of actions when the laws by which those actions are governed are autonomous with respect to the laws of Nature; that is to say, that they are self-willed and outside of Nature. In this way, the ego is distanced from the body and its physical circumstances.

This definition is typical of the idealist tradition, which relates free will to acausality (for example, Kant, 1788). The idea of free will in human beings is opposed to man-machine type French materialism, as in *L'homme machine* by La Mettrie (1749).
Other topics related to "freedom" will not be dealt with here. I shall speak, not of poetry or confused thought, but only of something that may be analyzed and whose meaning is understood. I shall not deal with sentiments, of the feeling of freedom that each one of us feels for himself. I shall instead concentrate on authors who speak explicitly of Nature rather than referring to the omnipotence of God. I shall eschew ethical/moral positions because these do not fall under the rubric of ontology. I discuss about being and not about what ought to be. Neither do I refer to freedom to enjoy certain rights, or to potential wishes, but rather to freedom to engage in the act of wishing. I should also stress that I do not concern myself with the question of being able to do what one wishes (freedom to exert one's will). The so-çalled compatibilists, when they do not fall into a non-explicit dualism (which would really be a kind of incompatibilism, for mind and body reside in two different realities) usually opt for defending these kinds of freedoms, which are "compatible" with a Laplacian Universe ruled by deterministic physical laws (Pérez Chico & López Corredoira, 2002). This is a very simple problem with a very simple solution: of course we are free in that sense unless something or someone hinders us from doing what we want (if we are in prison, for example). Any uncaged beast on earth has that freedom; even people manipulated by a sect has that kind of freedom because they can "do what they like." Likewise, any person manipulated by the communications media in our time of capitalism and democracy can be considered free in that sense because they may do, buy, and choose what they want and believe what they choose to believe. But is it not a little shocking to regard a puppet manipulated by the hand of fate as free? Hobbes (1654) asks:

> [H]orses, dogs, and other brute beasts, do demur oftentimes upon the way they are to take: the horse, retiring from some strange figure he sees, and coming on again to avoid the spur. And what else doth man that deliberateth, but one while proceed toward action, another while retire from it, as the hope of greater good draws, or the fear of greater evil drives him?

The question raised here is somewhat more profound: I ask whether one really wants what one wants ("freedom to want"), if the origin of the wanting is really mine or merely an effect of the laws of Nature; whether there exists an "ego" separate from Nature, or whether we are just Her puppets. That question has a less trivial solution and a subject referred to by many classical philosophers who have debated freedom; Hobbes (1654), Spinoza (1677), and Schopenhauer (1841) are prime examples. Those preferring a superficial, ingenuous view will be satisfied with the half-truths associated with the question of freedom of action. However, those who aim to reach the truth as closely as is possible with the human intellect, who would comprehend the hidden mechanisms behind appearances, who conceive philosophy as an exhaustive delving into things, will see that the problem deserving their attention is that of "freedom to want". There are those who believe that Santa Claus bring toys to children and are satisfied with that half-truth, which is not the complete truth until it is revealed that it is the parents, disguised as Santa Claus who actually fetch the toys. Likewise, some think that self-awareness bring us the gift of freedom of choice, but that is not entirely true; in reality, it is Nature, disguised as self-awareness, that is making the choices. The gifts may be enjoyed just the same, wherever they come from, but in the latter case we become adults who have cast off old myths.

Many modern authors may consider this notion of freedom somewhat *passé*, a relic of German idealism, and alien to present-day philosophy. However, it seems to me that we must not underestimate the importance of the influence on modern philosophy of Kant and his followers concerning the subject of free will. It will not do simply to dismiss the concept of freedom as merely the domain of idealists. In my view, it is a clear concept that expresses something that is not at first sight apparent and is contrary to the notion of "freedom to do". It would seem valid to criticize this definition of freedom linked to the concept of transcendental subject if, in exchange, we could provide an alternative clear definition that expressed what free will is, and that did not confound the question beyond comprehension as Schopenhauer (1841) describes:

> The question of freedom of will is in reality a touchstone upon which one may distinguish between those spirits who think deeply and those who limit themselves to superficialities, or again a crossroads where both part company, the former affirming the absolute necessity of producing motivated action of a specific character, whereas the latter maintain, along with the masses, freedom of the will. Then there exists an intermediate class, which, in a state of confusion, vacillates between both extremes, changing their own and others' aims, seeking refuge in words and phrases, confounding the question to the point of incomprehensibility.

It is not my intention with these words to deny the importance of the many works dealing with other concepts under the rubric of "freedom" throughout the history of philosophy, although I do insist in demonstrating that they focus on topics that are not of interest to comment on here. Since very many philosophers have used the word "freedom" and have dedicated many works to the subject, any attempt to give a global description of the discussions on freedom throughout the history of philosophy would lead us away from the purpose of this essay or from any other work of this length. Moreover, there exist several compilatory works that would be useful to those seeking an encyclopaedic knowledge of the subject that concerns us here: the work of Adler and coworkers (Adler, ed., 1973), for example, or the monumental work in four volumes by Vallejo Arbeláez (1980). I therefore limit the scope of this essay to the definition given above. I also offer two more reasons: 1) because I really believe that the subject of freedom is the one that corresponds to the said definition, and 2) because this acceptation of freedom is the one used by those who defend it within the framework of the contemporary natural sciences.

## 3. Ontological reductionism

Reductionism (or non-emergentism): The whole is the sum of its parts and their interaction. We must construe "is" in its usual sense of ontological being or existence. "Sum" is synonymous with union. "Parts" makes mention of an arbitrary partition or fragmentation provided one includes all its existing elements, even though this division need not be inherent in the being itself, it merely being convenient or seen from any subjective point of view. The ontological is the whole and its equivalence is with the sum of the parts and their interaction; the division is a mental representation that has no value in reality. New properties of the whole derive (they do not emerge) from the simplest interacting systems that comprise the laws that govern them.

Given that the level of partition referred to is unspecified, it is to be understood that it could be any. In other words, the whole is composed of a certain number of parts at a given level of partition, and those parts are in turn composed of subparts, etc. The whole might be the entire Universe, or we could equally consider just one of its parts –a star, for example— and proceed to apply the same reductionism. The transitive property of reductionism is also worth mentioning: if a system reduces to a number of parts and those parts reduce to subparts, then the system itself reduces to subparts.

The definition makes reference to ontology, that which the parts themselves are, and not to epistemology, what we know of things. It is thus that the term reductionism is to be understood in this section. Therefore, the term can be called "ontological reductionism" in order to differentiate it from other senses. Ontological reductionism is often confused with epistemological reductionism. Sometimes by reductionism is understood the use of the same terms in any scientific field, a tendency that should not be confused with that to which I refer here. Neither should it be confused with taking the qualitative properties of simple objects in physics as the sole element in science. Epistemological reductionism refers to the view that the branches of science are all particular cases of laws formulated by physics (Ayala, 1983), but nothing is said about reality in nature of this kind of reduction. For example, an epistemological reductionism would state that everything is a question for the physicists to deal with, whereas ontological reductionism would claim that everything is a question of physics, although not necessarily a problem to be dealt with by the physicists.

As has been pointed out by some authors (e.g. Oppenheim & Putnam, 1985; Ayala, 1983), one should be careful with this distinction among reductionisms. There is no doubt that all epistemological discussions of the methodology of distinct sciences provide much food for thought, but their clarification is not of interest here. Suffice it to say, in this essay I am interested in ontology, since we are addressing the question of the real existence or otherwise of the freedom of man, and are not the question of knowledge.

Ontological reductionism does not claim that, working with the most basic science, we may assess the behaviour of any system in the Universe, that from basic laws we may rebuild the entire Universe (Anderson, 1972). Biologists may continue to pursue their science while having no idea of the microscopic physical processes in play because their methodology is better suited to tackling problems at correspondingly higher levels of complexity. Every science undoubtedly has its categories (Bueno, 1995), and the categories of one science are not replaceable by those of another; but all this is affects only the epistemological level, gnoseological if you prefer, but does not affect the question of what nature is or the fact that all possible levels of nature are reducible to more basic levels. We shall also refer to sciences, but only because the natural sciences study Nature; and if we believe that what the sciences say in that regard is correct, we shall also be talking about Nature. In other words, we claim that everything is reduced to more basic laws, the laws that govern the most basic elements of Nature, the physical laws; those laws of physics that physicist study; thus the connection with the physical sciences. However, physicists as scientists do not study animal proteins, for example, even though these may theoretically be reduced to matter governed by the laws of physics; they do not do so because the methodology of their science and the categories that they employ do not allow them to tackle such a problem with the tools of a physicist because of the elevated complexity of the problem. In any case, the study of complex systems does not imply the need for new laws to substitute those of basic physics; it does, however, require distinct methods of analysis, but no new physical law not previously studied in simple systems (Godlenfeld & Kadanoff, 1999).

Clearly, a physicist cannot properly describe animal aetiology, for example, even though it can ultimately be physically reduced. He can, however, survey the fundamental properties of the animal in its environment (food, solar radiation, etc.) through the conservation of energy or the global increase of entropy (which the organism can reverse but at the expense of the environment). Calories are neither created nor destroyed (as many men and women with excess kilogrammes are aware); instead, the body ingests them in food, and they accumulate in the form of fat, or are expelled through physical exercise. And, as with the conservation of energy, so with any fundamental physical property present in any organism, however complex it may be. For example, the physical determinism or indeterminism that will be addressed in other chapters of this book must also prevail in complex systems, whatever their level of complexity. It is these more complex properties that cannot be derived from physics by virtue, I repeat, of their very complexity, and not because they lie beyond the reach of the physical laws of Nature.

As stated in a monograph in the journal *Science* on this question (Gallagher & Appenzeller, 1999), we have the very best of reasons for accepting the reductionist hypothesis: it works. Not that reductionism has been proved as one would prove a mathematical theorem; it is a hypothesis, not a corollary, but everything in science works within the framework of this idea. Any property of matter is explicable in physical terms. Organic chemistry is reducible to inorganic chemistry. Finally, or almost, in response to chemical vitalism, which opposed the reduction of living matter to inorganic chemistry, Berthelot, in the mid-XIXth century, provided the necessary chemical evidence to demonstrate that organic chemistry ultimately reduces to inorganic chemistry. The complete organism of living beings, or living beings in their totality, are reducible to a mass of living tissue, and the entirety of the matter in that tissue is expressible in terms of inorganic chemistry. That living beings are composed of organic structures that are themselves made up of living tissue, and that this tissue in turn comprises cells, have been well known in biology and medicine through the studies of anatomy and cellular biology for a long time. It is worthy of consideration, however, that, during second quarter of the XXth century, a decisive answer was given to the question of the reduction of biology to chemistry through advances in biochemistry, or molecular chemistry. The reduction of psychological phenomena is now being tackled by neuroscientists has had some success. Smith (1970) affirmed that neurobiologists have covered much ground towards a satisfactory physical theory of the living brain, and that there is now as little place for such strange and immaterial causes as mind within the machinery of this liquid state computer as there would be in the mechanisms of the computers used in industry to solve business problems.

We may summarize this section by saying that not everything is a question for specialists in the physical sciences, but everything is physical, everything follows physical laws.

## 4. Classical mechanics and determinism

For Newton, all physical systems could be reduced to a collection of *N* pointlike particles in space and their interactions. Any particle *i* possesses a given mass, $m_i$, and a position $\vec{r}_i(t)$ in three-dimensional space that varies continuously over time. Other kinematic properties of the particle, such as its velocity or acceleration, are merely the first and second derivatives with respect to time of the vector function $\vec{r}_i(t)$. After Newton, the electric charge of the particle was introduced into classical mechanics as another quantity associated with the particle; likewise, other characteristic properties of particles were also introduced, although these tend to fall under the purview of quantum mechanics rather than its classical counterpart.

In a closed system of *N* particles, the trajectories $\vec{r}_i(t)$ follow Newton's laws, which are expressed in mathematical form as follows:

$$m_i \frac{d^2 \vec{r}_i(t)}{dt^2} = \vec{F}_i\left(\vec{r}_j, \frac{d\vec{r}_j}{dt} \forall j \neq i; t\right) \forall i, \forall t, \quad (1)$$

where $\frac{d}{dt}$ and $\frac{d^2}{dt^2}$ respectively symbolize the first and second derivatives with respect to time of the function to which they are applied, and $\vec{F}_i\left(\vec{r}_j, \frac{d\vec{r}_j}{dt} \forall j \neq i; t\right)$ are force vectors, which are a function of the position and velocity of *j* other particles, distinct for the *i* particles, with respect to particles *i* and time *t*. The $\vec{F}_i$ components are exact numerical relations, so that, given the positions of the particles and their derivatives (the velocities), the values of the three components, $\vec{F}_i$, of the force affecting each particle will be determined for each instant of time. In his *Optics*, Newton considered that some forces active in Nature are, on the one hand, inertia and the impenetrability of bodies, and, on the other hand, a series of "active principles", among which he mentions gravity, electricity, magnetism, combustion, fermentation, chemical affinity, and even "vital forces". Such forces were indeed later reduced to a dependence on position and velocity.

There are *3N* second-order differential equations since there is a vector equation for each particle, which presupposes three scalar equations for each particle, given that vectors belong to a three-dimensional space. Such a system of equations as this has a unique solution provided that: 1) *6N* initial conditions are given (specifically, the six quantities given for position and velocity, $\left(\frac{d\vec{r}}{dt}\right)$ for each particle at a given instant; 2) the positions $\vec{r}_i$ possess first and second derivatives. The second condition always holds and, when the forces among the particles can be found –again, this condition is fulfilled where there are no divergences (infinities) in the expression for force.

If the interaction between the particles is gravitational, for example, infinities would arise if two particles occupied exactly the same position, their mutual distance being zero, and, the force being inversely proportional to distance, this would give rise to an infinite attraction. However, owing to the existence of mutually repulsive forces, such as the mutual electrostatic repulsion of the shells of neighbouring atoms, such infinitesimal approaches among particles do not occur, quite apart from the singularities that arise in general relativity.

In conclusion, one may infer that, given the positions and velocities of all the particles at a given instant of time in a closed system, their positions --and therefore their velocities and accelerations, derived with respect to time— will be determined for all instants in the past and future. The entire system for any value of *t* can also be determined for any other *6N* independent conditions –not linked to the differential equations— even though they do not correspond to the same initial time. On the basis of the foregoing discussion, it may be deduced that Newton's laws of motion imply that the future behaviour of a system of bodies is completely determined from the positions and velocities of the bodies at a single instant of time. Such was implied in the original Newtonian formulation, but it was Laplace who, more than a century later, drew attention to the determinism inherent in classical physics. The determinist theory grew in prestige during the XIXth century when areas of physics that did not at first seem to fit the determinist conception (thermodynamics, optics, electromagnetism) were finally reduced to the scheme of Newtonian equations (Fernández Rañada, 1982).

*4.1 Predictability is not the same as determinism*

When discussing determinism it is usual to see it mixed up with other quite distinct concepts, such as predictability and computability. "Predictability", or "computability", means that we human beings can predict the future state of a physical system, we can calculate the values of all its variables. It is a term that speaks of what we can know (epistemology), a matter that is somewhat different from ontological reference to determinism. It must be made clear that "determinism" is a wider-ranging concept than "predictability" or "computability". Determism does not imply predictability. Above all, we need to be clear that a deterministic system does not have to be knowable. There might be a destiny that determines an event, but knowledge of that destiny might be inaccessible to us; in other words, it might be unpredictable. What is certain, however, is that predictability implies determinism; that is to say, if we want to predict the behaviour of a system exactly, it must be governed by exact deterministic laws of which we have knowledge –along with all the parameters appertaining to it. It is one of the most important ideas of modernity that there exist laws, and that –thanks to science— we can know those laws.

Applying classical mechanics, we could predict the behaviour of a closed system if we knew the values of the positions and velocities of all its particles at a given instant of time and we could solve the system of equations (1). The fact that the system is determined does not imply that we can know its determination. When Laplace (1814) said, "An intelligence that knew...", he was not referring to human beings, for our knowledge will always be limited as the finite beings that we are. He was referring to determinism, and the mention he makes of an omnipredictive superintelligence is a mode of expression to enable him to refer to a predictability of the Universe that is theoretically possible, but denied to all non-infinite beings. In fact, he affirms that the possibility of attaining complete certainty is completely denied to man, and that the most to which he may aspire is to obtain a merely probably knowledge. In no case, at least as I interpret his lectures, may we understand Laplace's statement as the affirmation of our unlimited predictive capacity. He was well aware of the extremely high number of atoms contained in a few grammes of matter, and that knowledge of their positions and velocities far exceeds any human capacity.

Respecting human beings, the paragraph on superintelligence continues:

> [A]ll the efforts [of human beings] to seek truth tend merely to approximate it continuously to the intelligence that we have just posited, but from which we shall remain forever removed.

Laplace refers here not to predictability but to ontological determinism. Laplace's demon is a being of infinite capacity and can therefore know with absolute precision the position and velocity of each particle; he can make predictions to any degree of certainty. Predicting with infinite precision is equivalent to determining, and it is on this basis that Laplace with the term infinite predictability refers to determinism. However, human beings, far from being gods or all-powerful demons, do have a finite capacity, and our calculating power and the sensitiveness of the measuring apparatuses in our experiments and observations depend on our technological capacity, which is growing apace but is always limited. It cannot be infinite. Laplace adopts a very clear stance that does not admit of controversy: the world described by Newtonian mechanics is deterministic; the differential equations (1) have unique solutions once the initial conditions or environment have been established, and this implies determinism.

Any chaotic deterministic system, such as that arising in the problem of many bodies interacting gravitationally according to Newton's laws, is predictable up to a time that is constrained by the computational capacity of computers. To give a couple of examples of chaotic systems, we cannot with present-day computers, which carry out calculations over relatively short periods of time, predict what the weather will be beyond about a week, or in the positions of the planets in the solar system beyond a few million years.

Popper (1956), incorrectly in my opinion, put forward the view that classical physics does not require Laplacian determinism. He speaks of predictability (which he calls "scientific determinism"), and would have us believe that it is this form of determinism to which the majority of scientists and philosophers adhere to, and with which it makes sense for us to consider. The very use of the word "determinism" instead of "predictability" serves only to create confusion. One cannot say that classical mechanics is a correct model of the world and declare at the same time that there is no determinism. Prigogine (1994, 1996) and Prigogine & Stengers (1979, 1984, 1988) also resort to this kind of confusion.

*4.2 Crossroads*

The old idea of nineteenth century mathematician and physicist James Clerk Maxwell (Niven, ed., 1890) that the freewill of the individual acts in the form of crossroads (when the system can follow its destiny along one of two or more distinct paths), the tiniest fluctuations of a given variable of the system producing totally different future outcomes) arose with Prigogine. Following an argument based on thermodynamics, he alludes to indeterminism and the convergence of the Two Cultures (the humanities and the sciences) in relation to the concept of freedom. He declares that modern science approach a level of complexity that support indeterminism, independently of the new discoveries of quantum mechanics to which reference will be made in a latter section. But is it valid to proclaim a form of indeterminism in the behaviour of a system just because we cannot know the underlying determinations that lead to one or other form of behaviour when presented with two possibilities?

Reply: No. There might well be a cause that leads towards one path, however imperceptible at the macroscopic level (for example a simple atom pushing more towards one stable solution than to another), in the crossroads –and indeed there are such causes if we remain within classical mechanics. We are effectively dealing with unstable systems viewed macroscopically that can just as well choose one path as another through small variations, but those variations can, and indeed do, proceed in accordance with equation (1), following a strict determinism. Things are very different in quantum mechanics, as we shall see in section 5.

How does Prigogine respond to these accusations? He is undoubtedly aware of statistical mechanics and its implications. Why does he, then, radically insist on indeterminism? The answer he gives is that statistical description is irreducible to a set of microscopic variables, and that representation in the form of particles following their trajectories must be replace by a purely probabilistic description (Prigogine, 1994, p. 59). The probability distributions of a state do not physically characterize a large collection of particles, each with its own microstate. He therefore denies the very fundamentals of statistical mechanics and implies that thermodynamics is not reducible to classical physics. He believes that the macroscopic variables of the equation that governs the bifurcations is irreducible as a function of the positions and velocities of the components. He denies the bases of a firmly consolidated science whose results are corroborated in all areas except where quantum mechanics enters into play. He denies that thermodynamical quantities depend on what happens in the components of matter. He literally affirms that Boltzmann's attempt to explain irreversibility in terms of reversible laws is a failure (Prigogine, 1994, p. 41). The thesis held by Prigogine and his followers relies heavily on vague arguments in its favour. Prigogine aims to deduce from the dynamics of chaotic systems that the idea of trajectory should be abandoned and replaced by a probability theory of sets of possible trajectories in which these represent irreducible reality rather than representing our ignorance of the microscopic variables of the system. Classical statistics is fully consistent with the apparent paradoxes presented by irreversibility; there is no need for a change of ideas. This kind of conceptual revolution has in fact been set forth, but in terms of quantum mechanics, which we shall consider further on. However, as Batterman (1991) notes, in quantum mechanics the exclusion of hidden variables is supported by firm theoretical and experimental argumentation, whereas in Prigogine's proposal there reigns only unsubstantiated belief with no basis in rational argument to show that Boltzmann's conjectures were erroneous in this regard. Bifurcations correspond to chaotic systems, which are not predicable but are deterministic in classical mechanics.

*4.3 The arrow of time*

A temporal symmetry may be derived from Newton's equations (1). If we change the variable t in the equations to −t, then, given that $\frac{d^2\vec{r}_i(t)}{dt^2} = \frac{d^2\vec{r}_i(-t)}{d(-t)^2}$, the equations will remain unaltered. From the point of view of classical mechanics, there is no distinction between going forwards or backwards in time. This temporal symmetry is linked, among other things, to determinism –a numerically exact causality between the past and the future— by means of which it is possible to determine the past from the future and similarly the future from the past. All destiny is written in a single time: given the past, present, or future, the states of the system for any other time can be determined. We could even say that the passage of time does not exist, that there exists only a Universe that is always the same.

However, we do indeed perceive the passing of time, and we sense the notion that time has a direction, the "arrow of time", and it is this sensation that led many scientists and thinkers of our time to oppose the conception of temporal symmetry. Both Popper (1956) and Prigogine (1996), as well as Prigogine & Stengers (1979, 1988) defend the arrow of time and the indeterminism of classical physics. What is more, they base themselves on the former in order to demonstrate the latter. According to these authors, once the reality of time has been established, the greatest obstacle to achieving a greater rapprochement between the humanities and the sciences would have been eliminated. We no longer have a reason for choosing between practical freedom and a purely theoretical determinism. Mention is made of the arrow of time in the second law of thermodynamics, according to which the entropy, *S*, of the Universe, a measure of its disorder, is increasing. This defines an asymmetry in time: entropy increases towards the future and decreases towards the past. To give an illustrative example, this law tells us that heat flows from hot to cold bodies, or that we can mix fluids merely by putting them into contact with one another, but we cannot then unmix them. In these two examples, we see how irreversibility arises: we cannot make heat pass from cold to hot bodies, or cause mixed fluids to unmix.

Reply: There is no irreversibility in the sense of impossibility; the only reason that we see that some phenomena that suggest an arrow of time, such as why we do not see heat flowing from cold to hot regions, is that such a phenomenon is highly improbable, but not impossible. The second law of thermodynamics explains the reason for this in terms of the deterministic motions of the atoms that make up a physical system and in no way implies a contradiction. In other words, the arrow of time cannot be used to refute determinism. Time is irreversible in unstable systems because the behaviour of such systems in an inversion of the arrow of time is chaotic: very stringent conditions of low probability are required to render it possible to recover a past state in an irreversible process, and small variations in those conditions put beyond our reach any possibility of achieving reversibility. Given that space is continuous, the number of possible microstates in a configuration of particles tends to infinity, and only relatively few possess suitable initial conditions to be able to recover the past state of a system in a temporal inversion when the system is not in equilibrium. Prigogine, however, insists that this arrow of time implies indeterminism because he discounts fluctuations due to entropy; instead, he postulates that reversibility is absolutely impossible rather than merely improbable. His idea is that physical states evolve from non-equilibrium to equilibrium, and that the former are singular states that have all probability concentrated in a multiparametric space of zero size, whereas states in equilibrium can be non-singular. At the moment when one microstate leaves the state of non-equilibrium it ceases to be in a singular state and it is improbable that it should return to that state because of its zero size in the parameter space that allows zero probability for its return. But, as Sklar (1992, "The problem of Initial Probability Distributions") points out, such singular states are unnecessary and represent poorly real physical situations of interest.

It is perfectly possible to explain irreversible phenomena within the fundamental laws of classical mechanics. This was done by Boltzmann (1905) more than a century ago, and his arguments still hold good today:

> Given that in the differential equations of mechanics there does not exist any analogue for the second law of thermodynamics, this last cannot be mechanically represented other than by means of suppositions that conform to the initial conditions

The initial conditions of a system, that is to say, the positions and velocities of each of its particles, contain information about the evolution of the system in one sense or another, not the actual differential equations, which, as we well know, are symmetrical with respect to time.

## 5. Quantum mechanics and free will: counterarguments

Quantum mechanics is not metaphysics and we might therefore think that the subject of free will is beyond its analysis. I am partly in agreement with this position and indeed in this section I shall not try to put forward any metaphysical speculation, but rather the contrary: I shall try to refute speculative ideas that are often present in quantum mechanics. I agree with Physics Nobel laureate Richard Feynman when he says, "If we have an atom that is in an excited state and so is going to emit a photon, we cannot say when it will emit the photon. It has a certain amplitude to emit the photon at any time, and we can predict only a probability for emission; we cannot predict the future exactly.

This has given rise to all kinds of nonsense and questions on the meanings of freedom of will, and of the idea that the world is uncertain" (Feynman et al., 1965).

*5.1 Indeterminism in quantum mechanics*

Discussion on free will usually begins with consideration of the possibility of ontological indeterminism. Classical physics is a deterministic model of the world. We can speak of unpredictability but not of indeterminism in the laws of Nature. However, the orthodox interpretation of quantum mechanics accepts indeterminism in observables once they have been measured. For example, we do not know when an atomic nucleus is going to disintegrate. In the wave function $\psi$ representing a particle or system, the evolution is deterministic in any interval between measurements. The Schrödinger equation before the measurement is made is of first order with respect to time; therefore, given an initial state $\psi(r,t_0)$ at a fixed time $t_0$, the function $\psi(r,t)$ will be determined for any position $r$ and any time $t$. There is no indeterminism in the interval between two measurements; such indeterminism appears only when a measurement is carried out (see, for example, Cohen-Tannoudji et al., 1977). There are other interpretations that do not require indeterminism in their formulation, such as that of Bohm (1952), but the most extensive interpretation of quantum mechanics accepts ontological indeterminism in its formulation.

This indeterminism is present when quantum effects are important and weakens in large collections of particles (the macroscopic state), converging to Newtonian mechanics, so long as there exists a mechanism in which the state of a macroscopic system depends on what happens with a few microscopic particles.

How can a macroscopic motion depend on what happens with only a few molecules? The answer to this question is provided, for example, by investigating the application of quantum mechanics to the biology of human, or any other kind of living, beings; more specifically, in terms of certain studies within neuroscience that take quantum effects into account. Lillie (1927) draws attention to the possible implications of quantum indeterminism in macroscopic biological systems that are distinguishable from macroscopic systems of macroscopic components, such as wind up clocks, whose indeterminism does not transmit to lower scales of the system.

Various mechanisms have been proposed that could transfer this indeterminism to the entire human brain, and therefrom to the entire body, for the nervous system regulates the muscular movement of the entire living being. To be specific, it is believed that that is what happens in the phenomenon known as "synapsis", which consists in the exchange of neurotransmitters:

- One such mechanism alludes to the presynaptic membrane at the tips of the axon being a lipid bilayer, a layer two molecules thick, with one molecule for each layer. The entry or exit of the neurotransmitters in the neuron strongly depends on what happens in this bimolecular layer, which serves the function of master switch (Scott, 1985). The possession a thickness of only two molecules manifests the microscopic effects of the bilayer; whether or not any neurotransmitters pass, giving rise to movements of the body, is dependent in an indeterminate manner on what happens in the bilayer. The amplification of indeterministic behaviour is therefore made possible through such a mechanism as the one just described. Eccles (1973, 1975, 1994) and Beck & Eccles (1992) support this theory.

- Microtubules, protein molecules in dendrites and axons, are compound aggregates of particles that can be in two distinct states, depending only on the position of an electron. Microtubules take part in the control of the synapse so that the state of the brain depends on these molecules (Mitchison & Kischner, 1984a, 1984b; Penrose 1994; Rosu, 1997).

*5.2 Interaction of body and mind*

Given that the ideas of determinism and the negation of freedom traditionally come together, there has been an attempt to see in quantum indeterminism a distancing from classical mechanics that allows freedom (Eddington, 1932; Jordan, 1944; Frank, 1957; Margenau, 1961; Stapp,1995).

If indeterminism affects the macroscopic behaviour of biological systems, it will also apply to human beings. On the strength of this argument Jordan states:

> If the supposition is correct that the controlling reactions of organisms are of atomic physical fineness, it is evident according to our modern knowledge that the organism is quite different from a machine and that its living reactions possess an element of fundamental incalculability and unpredictability. One can object that our fundamental understanding of life phenomena is not greatly aided by considering a statistically functioning dice cup instead of a machine as the pattern of organism. But at the moment it is only important for us to determine in the negative sense that the machine theory of organisms (including their further results; e.g., in the sense of a denial of the freedom of the will) can hardly exist in view of the new physics. (Jordan, Engl. Vers. 1944)

> We can now know the behavior of an individual organism--regardless of whether animal or man--is not exclusively determined by mechanical necessity; we can no longer, with LaMettrie, forbid the soul or the will to intervene in the fixed and predetermined movements of the body's atoms. (Jordan, Engl. vers. 1955)

Among existing interpretations of quantum mechanics, those that give rise to control by a free being over the matter that constitutes the physical world are based on notion that the mind collapses the wave function during the process of measurement. Among the pioneers of this idea are Compton (1935, 1981), von Neumann (1932), and Wigner (1961, 1967), and it is supported by such authors as Stapp (1991, 1993, 1995), Bass (1975), Heitler (1963), Marcer (1992), and Penrose (1994).

In reality the idea is not a new one. This defence of free will is extraordinarily similar to the ideas of the epicureans (for example, Lucretius' *De rerum natura*), who speculated that the atoms of the human body could change their trajectories according to the will seated in the mind of the body's owner. Indeterminism leaves open the possibility of choosing between diverse solutions in the trajectory of atoms, and it is man's will that chooses it. In the present quantum version, human will governs the body by means of the presynaptic membranes, or more specifically by the microtubules. The synaptic connections are controlled by the mind, and the system of neurons would be continuously influenced by them through the activity of the citoskeletons, thus giving rise to "free will" (Penrose, 1994; Horgan, 1994; Rosu, 1997).

There are two aspects to the hypothesis that the mind interacts with the body:

- The measuring apparatuses cannot produce the collapse of the wave functions; they couple (in a superposition of states) to the system observed. Thence is deduced the need of another class of substance that produces the collapse of the wave function: the mind (von Neumann, 1932; Wigner, 1961, 1967). Therefore, the mind can choose the state of all the systems that it observes. That includes the neurotransmission in the brain associated with the mind.
- The mind coordinates all the neuronal states when it generates a thought. This explains how a property of quantum coherence in the brain, the nonlocality in all the particles confers to the brain a unity within a unique quantum state (for example, Stapp, 1991; Josephson & Pallikari-Viras, 1991).

This is clearly a dualism, an interpretation split off from traditional scientific materialism. It is rather an interactionist dualism: the presynaptic membranes, in other words the microtubules, play a role similar to that of the pineal gland that connects the body and soul in the dualism of Descartes (1641). In this perspective are in some way included the ideas of Popper, who, however, localizes the region of interaction with more exactitude; that region does not embrace all cerebral synaptic processes. Together with John Eccles, a neurophysiologist, he proposes an argument for interactionism by defending the existence of a place in the left hemisphere of the brain where mind and brain interact (Popper & Eccles, 1977). In their thick volume, both Popper and Eccles defend freedom with a position allied to quantum liberty; in other words, the mind observes the brain and selects the neurons in order to activate them with the objective of attaining what it wants, of obtaining the mental outcome it desires. In reality not all the authors who speak of freedom within the framework of quantum mechanics declare themselves to be interactionist dualists as does Popper, instead some speak of a mental world "emerging from the physical world". Wigner is uncomfortable with being denominated a dualist and proposes a freedom derived from quantum mechanics that counters the arguments of Popper and Eccles. "No. There is not only a world!", this theoretical physicist asserts. He believes that the new laws of Nature involve human consciousness in the only existing world. But this monistic attempt is a hidden form of dualism.

*5.3 Counterarguments (I): indeterminism does not imply free will*

There are two options: 1) determinism; 2) indeterminism. If case 1) holds, then the atoms in our body follow strictly deterministic laws, and no intervention by us is possible; we cannot be the first cause of the movements of the atoms in our body but are limited by a Laplacian determinism that leaves no space for free will (that is, for free will as defined in section 2, and not for popular and naive notions such as doing what one wants). The interesting case here is the second, that of indeterminism. Let us suppose that quantum mechanics provides a correct theoretical framework within which to put forward an ontological indeterminism (not merely unpredictability as in classical mechanics). What happens to free will in this case? There could be free will, but not necessarily because "indeterminism does not imply free will". Clearly, indeterminism is a necessary condition for free will but it is not sufficient. The libertarian tradition had for centuries been mulling over the subject of determinism when they were asked about freedom, devoting so much time to opposing determinism, that it has seemed as given that the opposite to determinism is freedom. At least that is how some authors have seemed to assimilate the matter, those authors who quickly applauded quantum indeterminism on the grounds that thus could be found freedom.

It is erroneous to think that the discussion on free will is necessarily a discussion concerning necessity or contingency. As Arana (2005) sustains, in the present debate freedom is not to be found between necessity and chance but instead must find its place in opposition to chance and necessity. I believe that there indeed lies the crux of the question: there cannot be freedom while there is only chance or necessity. As Kant (Op. Post. 1928, refl. 5369) pointed out, freedom is neither Nature nor chance. Those philosophers or scientists who hold that indeterminism is free will forget the rules of classical logic and tell themselves: $p\rightarrow\neg q \Rightarrow \neg p \rightarrow q$, where "p" is determinism and "q" is free will. Such a deduction is incorrect.

Think, for example, of a robot that follows indeterministic laws of chance. It is free? No it is not. Indeterminism is not the absence of causation but the presence of indeterministic causal processes (Fetzer, 1988). I mean that "causality" is not necessarily determinism; we may understand "causality" in a wider sense: causality as "explanation" or "reason". An explanation or reason for an event means the following of a law (a law of statistics, perhaps), and the presence of laws is the absence of free will. Quantum mechanics is non-deterministic but it is not acausal. There is always a cause, an explanation or reason, for any phenomenon. For example, when an electron is pushed towards another electron both are repelled and their positions and velocities are indeterminate. The cause of the repulsion are the laws of electrostatics. The electrons are not free to choose whether or not to be repelled.

Renouncing the fatalism of scientific materialism requires disabusing oneself of any idea of causality, freeing oneself, therefore, from any explanation. The moment an act, a choice, etc., can be explained in terms of physical laws, even if probabilistic, we are including that action, choice, etc., as another phenomenon of Nature and hence negating that its origin is in ourselves, like something that is independent of Nature.

*5.4 Counterarguments (II): scientific knowledge of consciousness*

An argument in favour of mind based on physics cannot avoid being in a sense gratuitous; it is an opinion rather than an empirical or rational result that is proper to science.

Wigner and von Neumann point out that consciousness is necessary for an understanding of quantum mechanics in order not to fall into contradictions, but their reasoning has a number of tremendous flaws. Their differentiation between mutually coupled systems and those which, on merging, produce a collapse of the wave function, giving rise to a measurement, seems reasonable enough. Their observation that human beings make measurements seems to be correct. The doubtful inference is the attribution of a collapsing character to something that is possessed exclusively by human beings, namely mind. The fact that measurements are associated with collapse, and that human are present during the measurements does not imply that humans are responsible for the collapse. To say that human beings are present during the measurements is tantamount to saying nothing, given that humans have knowledge through their interaction with the known object. If there were no humans present they would have no knowledge and would make no measurements.

The answer that might be expected is: if the human mind does not produce the collapse during the measurement, what does produce the collapse? Is there an alternative to the hypothesis that mind collapses wave functions? Yes, it is clearly possible to conceive a quantum theory without invoking consciousness (Stenger, 1997), a materialist and reductionist position is quite consistent with observations and quantum theory. The role of the human observer in a measurement does not have to possess any particular significance and can even be considered unnecessary (for example, Shimony, 1988; Mulhauser, 1995). There certainly is such an alternative. For example, the physical system itself, which includes both the measuring apparatus and the system to be measured, macroscopically produces the collapse on itself of the wave functions associated with its microscopic components.

In effect, the interpretation of Bohr, along with most of the greatest specialists in quantum physics, is that it is not consciousness but the distinction between the investigative apparatus and what is investigated that is the central element of the act of fundamental quantum observation. A machine with a computer can do the experiments and an observer read off the results provided by the machine once the result has been recorded in its memory. This cannot be verified given that we cannot know whether the collapse has been produced by the computer and measuring apparatus or by the human mind when the observer checks the result, but we do at least know that quantum mechanics can be interpreted without the notion of autonomous mind in the role of the observer.

Neither should coherence of all the parts of a quantum system, whose states are joined and interdependent until the collapse is produced, be confused with a coordinating direction of the system, like a conductor with his orchestra. The parts of a physical system can indeed be compared with the parts of an orchestra in which the musicians listen to one another's playing, but there could be notable cohesion only at low energy levels (at low temperatures), or where there hardly existed any interaction with the surroundings (Mulhauser, 1995). Even in the case that it was discovered at some time in the future that the human brain could maintain such special conditions, like superconductors at low temperature, the orchestra conductor is still missing, and neither classical nor quantum physics can have anything to say on the matter. The arguments of Stapp (1991), which differentiate between classical and quantum mechanics, cannot alone constitute an argument on the mind-body problem.

It has to be concluded that contemporary physics has not brought us any nearer to a knowledge of an autonomous consciousness "freely" governing the body. The introduction of the new concepts of quantum mechanics is irrelevant to the mind-body problem, as is explained in an article by Ludwig (1995), "Why is the difference between quantum and classical physics irrelevant to the mind/body problem?" We could adopt a subjectivist stance, just as at the time of the predominance of Newtonian mechanics one could be a Berkeleian idealist, but such points of view are nothing more than opinions beyond the scope of science.

Physicists have nothing to say regarding mind, but neuroscientist most certainly do. Nowadays, most neuroscientists believe that the idea of an autonomous mind or soul is a myth. Instead they adopt a materialist position in which mind can be explained in terms of neurological processes (Crick, 1994).

There are many examples illustrating materialist philosophy in the neurological sciences. The electric signals applied to the brain produce variations in our consciousness, such as images and memories (Penfield & Perot, 1963). The opposite is also true: electrical signals are registered in the brain that are associated with any conscious sensation. Moreover, electroencephalograms register activity before the individual is "conscious" of the sensation. For example, in an experiment (Deeke et al., 1976; Libet, 1985, 1987) in which a number of volunteers connected to encephalographs moved their fingers at will the apparatus registered electrical activity lasting about 300 milliseconds before the person became conscious of taking the decision to move the finger. This clearly implies that the unconscious activities of the brain fires off the activity of the neurons, and this means that matter governs consciousness before consciousness governs matter. The mind cannot be autonomous in the taking of decisions. Furthermore, it is clear that a minimum time (60–70 seconds) is required for the brain to carry out operations that the simplest conscious sensations produce (Libet, 1985). This time interval is too long for it to be considered a spontaneous decision. Neurons cannot wait so long for instructions from the mind in order to execute a decision on the state of a neurotransmission.

The quantum coherence state property of the entire brain is also absurd if we bear in mind certain experiments in which the neuronal unity between both cerebral hemispheres is severed. In this case, two wills are made manifest (Sperry, 1964). The unity of the brain is owed to "neurological connections" and not to any quantum coherence property. Neither can we relate the will (our capacity to take decisions) with the capacity to observe during a measurement, as proposed by quantum subjectivism. There are areas of the brain (Brodmann's area 24) directly related to will and, when these are damaged, the patient can observe (measure) but the brain cannot send an order for the body to move (Crick, 1994, post scriptum).

All these examples constitute experimental evidence, and not mere hypotheses based solely on speculation. And all these facts indicate that there is no autonomy in the brain, no "ego" sending orders to the body. The defence of free will is impossible in this context, and speculation on the subjective role of consciousness in the measurement of quantum systems is absurd.

*5.5 Counterarguments (III): evolution and ontogeny*

The problem that most philosophers who support an origin of mind in the context of the theory of evolution have is that they are either insufficiently familiar with the theory or forget many of its important points. In general, they do not even concern themselves with the origin of mind. It is there –so they think— and has somehow emerged from matter in some way; they think that the word "emergence" resolves all the problems and enables them to speak of the spiritualization of matter. The same happens when we ask ourselves about the origin of mind during the growth of a human body through cellular multiplication (ontogeny) from the moment in which the spermatozoon and ovum unite. Leaving aside religious and personal beliefs, we must consider that neither evolution nor ontogeny can explain the emergence of an autonomous mind that collapses wave functions in matter. The DNA copy contains only instructions for the building of proteins in various "material" tissues. Mutations are merely alterations of DNA. And on goes the argument. It is useless insistently to intone, "Emergence, emergence!" while we are in a world of "matter, matter!" Emergence of what? Of a mind that distinguishes matter while at the same time itself being generated by matter? Absurd.

According to quantum subjectivism, mind should emerge spontaneously. It is not possible to speak of a half or a quarter of a mind producing a half or a quarter of a collapse, or that there is no mind and matter couples with other physical systems. This is the position of von Neumann, Wigner, and others. We must therefore think of the spontaneous creation of mind. Consider a baby, a human cub, without mind and consisting purely of matter. A second later we have a baby with a mind that can produce a collapse of the wave function of the systems it observes. Absurd.

The problem most difficult to solve is the paradox of the Universe before the existence of any mind (Bohm & Hiley, 1993). If mind produced the collapse of wave functions in matter, then Nature prior to the existence of mind would be without such collapse, and the real Universe could not have been born because it would have to be in a superposition of states. Absurd. Even more ridiculous ideas have been put forward to explain this paradox (for example, Kafatos & Nadeau, 1990): a universal Mind (God?)

was present before the existence of life on Earth and produced the collapse of the wave functions. This pantheistic solution does not explain why human minds are now responsible for that collapse rather than the Mind of God. Did God's mind vanish after the appearance of our mind on Earth? Absurd.

## 6. Conclusion

I briefly summarize below the main conclusions of this essay:
- The concept of freedom that is worth discussing for its non-trivial metaphysical content is the freedom of man with respect to the laws of Nature and the origin of his volitions.
- The positions of ontological antireductionism are just opinions outside the remit of science and existing empirical evidence.
- Indeterminism is a necessary but insufficient condition for free will to exist. Spontaneity is not the same thing as chance.
- The description of the world derived from classical physics is deterministic.
- The description of the world derived from quantum physics does not have a unique interpretation, but it is difficult to see how indeterminism is refutable.
- The dualistic role of mind in quantum mechanics, essential if free will is to be deduced therefrom, comes into conflict with many experimental findings of neuroscience.
- From the point of view of the evolution of species, it quite impossible to put forward a non-materialist theory for the creation of mind that quantum libertarians proclaim as reality.
- The position of the natural sciences with regard to free will continues in our time to resemble that of the French materialist, except in decontextualized postures that use scientific terminology to refer to sundry speculations that have no grounding in objective terms.

All the preceding argumentation would be unnecessary if we were to admit what seems to me and others seems utterly trivial: science, dealing as it does with what is objective, cannot defend the idea of freedom, which requires autonomous recognition of the subjective. The development of the argument given here is in a sense a tautology regarding the simple fact of the determination of some scientists and thinkers to deny it. Science –past, present, and future— can never defend the hypothesis of the freedom of man. It is no longer a question of enter into a detailed discussion of quantum mechanics; neither is it a question of waiting for a new theory to provide a suitable defence. It is simply that science and freedom cannot fit into the same holdall. Libertarianism must follow a path that carries it far from science.

Most thinkers have difficulties with the justification of responsibility without the concept of freedom. Fearing a social disorder in which one can licitly cause harm by alleging lack of responsibility and blaming either Nature or society for any ensuing ill, these thinkers appeal to the "human dignity" and the need to defend free will or freedom at all costs. In spite of the common use of the expression "human dignity", it nonetheless remains a poetic allegory rather than a term based on rationality. Skinner (1971) considers, quite reasonably, that the use of phrases such as human "freedom" and "dignity" derives from detestable superstitions. They who proclaim the need for freedom at all costs are in reality men of little faith who lack the human capacity to live in the desert of nihilism. We probably cannot continue to apply the term "responsibility", but we can establish some criteria of social order, punishment specifically, that is directly incumbent on the said order with the aim of avoiding acts that are prejudicial to a community. It is not necessary to be responsible (in the sense of having chosen "freely") in order to receive a punishment. The punishment can be applied for other reasons, such as the prevention of future crimes or misdemeanours.

Neither is it the case that poetic allegories escape the sensibilities of those who adopt a fatalistic point of view. As Boltzmann (1903) said (and I agree with him):

> I can just imagine the horror that my thesis would inspire in the idealist, who would fear to see swept away all that is grand and sublime, and all that is poetic ruined in favour of a mechanism that is devoid of all sentiment. But such a fear would seem to me to be based on a complete misapprehension of all that I have said thus far.

Effectively, there is sublime poetry in the idea that everything can be reduced to mechanical materialism, and there is a certain beauty in the idea. Far removed from science, in philosophically inspired literary prose and verses, we also find many examples of the aesthetic exaltation of fatalism. Love of fatalism is considered by some great poets to be the greatest glory of human understanding. For example, Omar Khayyám in the *Rubáiyát* poetically describes the passionate resignation of an absolute religious determinism. This philosopher-astronomer-poet of the XI/XII century meditates dolorously on the pain of existence and the death that surely awaits us.

The denial of free will that will no doubt continue to undermine many past and future philosophical systems, does not mark the demise of thought; neither is it the destruction of human dreams. Quite the contrary. It marks the beginning, the beginning of thinking that raises man beyond his dreams. It is the road to unlimited creation. It is the eternal dream that mixes life and death, being and not being. As human beings we shall attain our desired immortality.